\documentclass{pasj00}

\begin{document}

\title{Spatial Distribution of the Galactic Center Diffuse X-Rays 
and the Spectra of the Brightest 6.4~keV Clumps}

\author{Katsuji \textsc{Koyama}, 
Yojiro \textsc{Takikawa}, 
Yoshiaki \textsc{Hyodo}, 
Tatsuya \textsc{Inui},  
Masayoshi \textsc{Nobukawa}, 
Hironori \textsc{Matsumoto} and Takeshi Go \textsc{Tsuru}}

\affil{Department of Physics, Graduate school of Science, Kyoto University, Sakyo-ku, Kyoto 606-8502}

\email{koyama@cr.scphys.kyoto-u.ac.jp}


\maketitle

\begin{abstract}

The high energy resolution and low background, particularly
in the hard X-ray band, of the X-ray Imaging Spectrometer
onboard Suzaku provide excellent spectra of the Galactic
center diffuse X-rays (GCDX). This paper reports on the
results of spatially resolved spectroscopy of GCDX.  The
most pronounced features of GCDX are the K-shell transition
lines from neutral (Fe\emissiontype{I}) and He-like
(Fe\emissiontype{XXV}) irons at energies of 6.4~keV and
6.7~keV, respectively. The fluxes of these lines are
non-uniformly and asymmetrically distributed with respect to
Sgr A*. The 6.4~keV lines are particularly bright on the
positive side of the Galactic longitude (east-side) with
clumpy structures.  A bright clump near the GC exhibits a
time variability over a timescale of a few years.  Neither
the 6.4~keV nor 6.7~keV line flux shows close
proportionality to the continuum flux (5--10~keV band); the
6.4~keV line shows excess on the high flux side, and vice
versa for the 6.7~keV line.  On the other hand, the sum of
the 6.4~keV plus 6.7~keV line fluxes with a ratio of 1:2
shows good proportionality to the continuum flux, and hence
we phenomenologically decomposed the continuum flux of the
GCDX into the 6.4~keV- and 6.7~keV-associated continuums
with a flux ratio of 1:2. Based on these facts, we have
tried to estimate the contribution of diffuse and integrated
flux of point sources to the GCDX.

\end{abstract}

\section{Introduction}

The Galactic center diffuse X-rays (GCDX) exhibit many
K-shell lines from highly ionized atoms, such as the
K$\alpha$, K$\beta$ and K$\gamma$ lines of He-like
(Fe\emissiontype{XXV}) and H-like (Fe\emissiontype{XXVI})
iron, and the K$\alpha$ line from He-like nickel
(Ni\emissiontype{XXVII}) (e.g. \cite{Ko07d}).  These K-shell
lines carry key information about the plasma physics. The
electron and ionization temperatures of a hot plasma are
constrained by the line flux ratio of
Fe\emissiontype{XXV}-K$\beta$ to
Fe\emissiontype{XXV}-K$\alpha$ and that of
Fe\emissiontype{XXVI}-K$\alpha$ to
Fe\emissiontype{XXV}-K$\alpha$, respectively.
\citet{Ko07d}, using these line ratios, reported that the 5--11.5~keV 
band spectrum of the GCDX is naturally explained by a
6.5~keV-temperature plasma in collisional ionization
equilibrium (CIE) plus a power-law component with a photon
index of $\mit\Gamma$=1.4.  The former component has nearly
the same flux as that from the latter component (see figure
7, in \cite{Ko07d}).  The origins of these components and
highly ionized atomic lines are, however, open questions.

In addition to these highly ionized atomic lines, the
K$\alpha$ and K$\beta$ lines from neutral irons
(Fe\emissiontype{I}) and the K$\alpha$ line from neutral
nickels (Ni\emissiontype{I}) have also been discovered. A
likely origin of these neutral K-shell lines is due to
fluorescence irradiated by external X-ray sources
(e.g. \cite{Ko96,Mu00,Mu01,Ko07b,Ko08,No08}).
However, alternative scenarios, such as a bombarding of
energetic electrons, have also been proposed
(e.g. \cite{Pr03,Wan06,Yu07}).

The key questions are: what are the origins of the 6.7~keV
and 6.4~keV lines, and how much is the contribution of
unresolved point sources?  The 6.7~keV and 6.4~keV lines in
the GCDX are spatially and spectrally entangled with each
other.  Spatially resolved spectroscopy could disentangle
this complicated situation.  Together with good energy
resolution and low background near and above the $\sim$6~keV
band, Suzaku (\cite{Mi07}) is the best satellite to date for
this study.  This paper focuses on the 5--11.5~keV band
X-rays in the sub-degree region near the Galactic center
(GC). The distance to GC is assumed to be 8~kpc
(\cite{Re93}).  In this paper, quoted errors are at the 90\%
confidence level, unless otherwise mentioned. We use the
Galactic coordinates; hence, the east means the positive
Galactic longitude side and vice versa for the west.

\section{Observations and Data Reduction}

Two pointing observations (here, the east and west fields)
towards GC were performed in September of 2005, with the
X-ray Imaging Spectrometers (XIS; \cite{Ko07a}), at the
focal planes of the X-Ray Telescopes (XRT; \cite{Se07})
onboard the Suzaku satellite (\cite{Mi07}).  The
data-selection criteria and subtraction method of the non
X-ray background (NXBG) are the same as those given in
\citet{Ko07d}. The charge transfer inefficiency (CTI) and
fine gain-tuning of CCD-to-CCD and segment-to-segment levels
(for the CTI and the CCD segment, see \cite{Ko07a}) were
self-calibrated using the K$\alpha$ lines of
Fe\emissiontype{XXV} (6.7~keV), Fe\emissiontype{I} (6.4~keV)
and Helium-like sulfur (S\emissiontype{XV}) (2.46~keV). The
absolute gain-tuning was made using the $^{55}$Fe
calibration sources irradiating the CCD corners, and also
using the Fe\emissiontype{XXVI}-K$\alpha$ line, which has
a relatively simple structure. The over-all systematic error
of our gain determination near the iron and nickel K-shell
energies is estimated to be within $^{+3}_{-6}$ eV.  The
details concerning the calibration procedures and the results are
given in \citet{Ko07d}.

For a timing study of the 6.4~keV line flux, we used the
Chandra GC data obtained by the Advanced CCD Imaging
Spectrometer array (ACIS-I) with total exposure times of
$\sim$500~ks. The logs are ObsID 2943, 2951, 2952, 2953,
2954, 3392, 3393, 3663 and 3665, observed in 2002.  The data
were reprocessed, using the CIAO version 3.4 and the
calibration database version 3.4.0.

\begin{figure} 
  \begin{center}
    \FigureFile(80mm,60mm){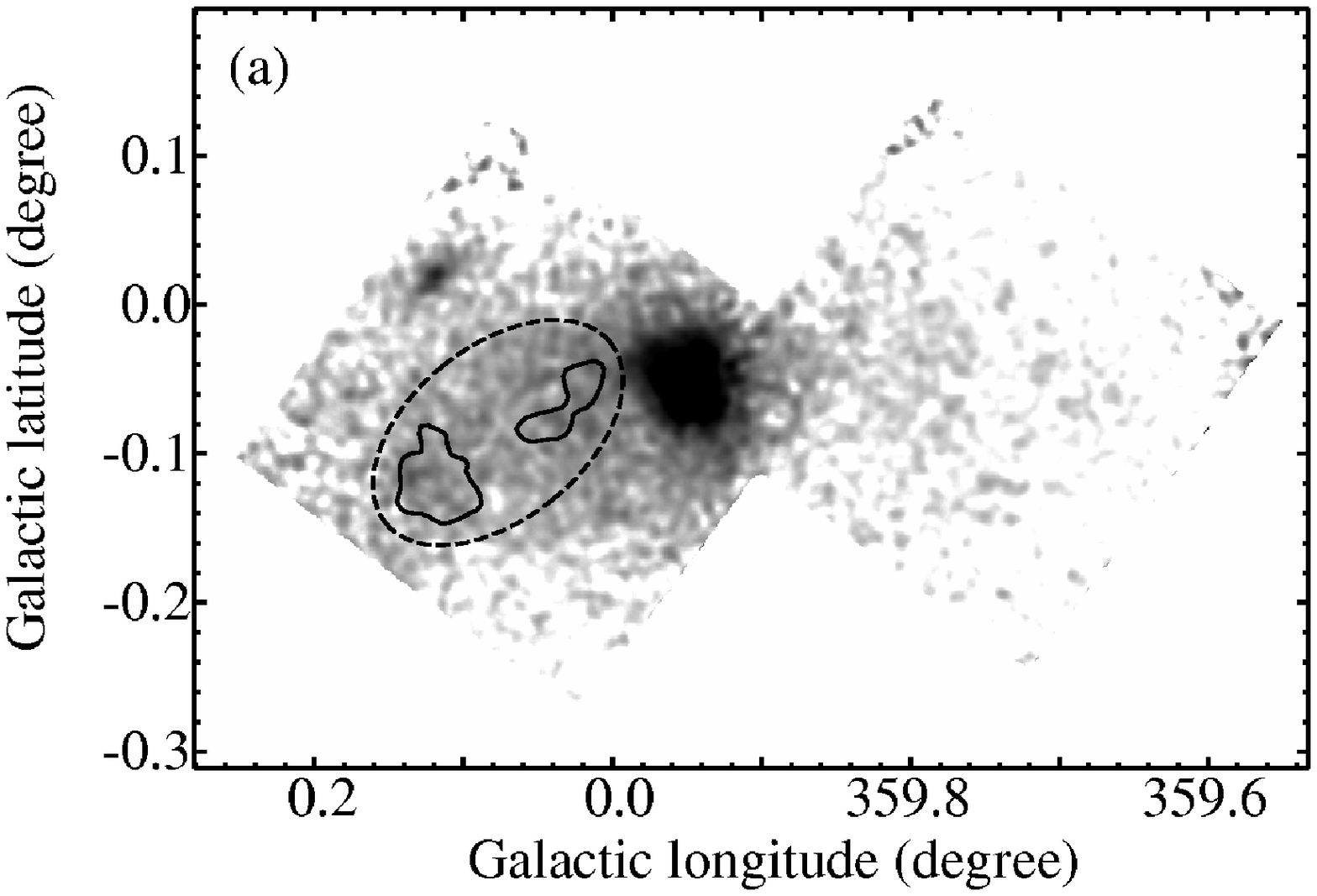}
    \FigureFile(80mm,60mm){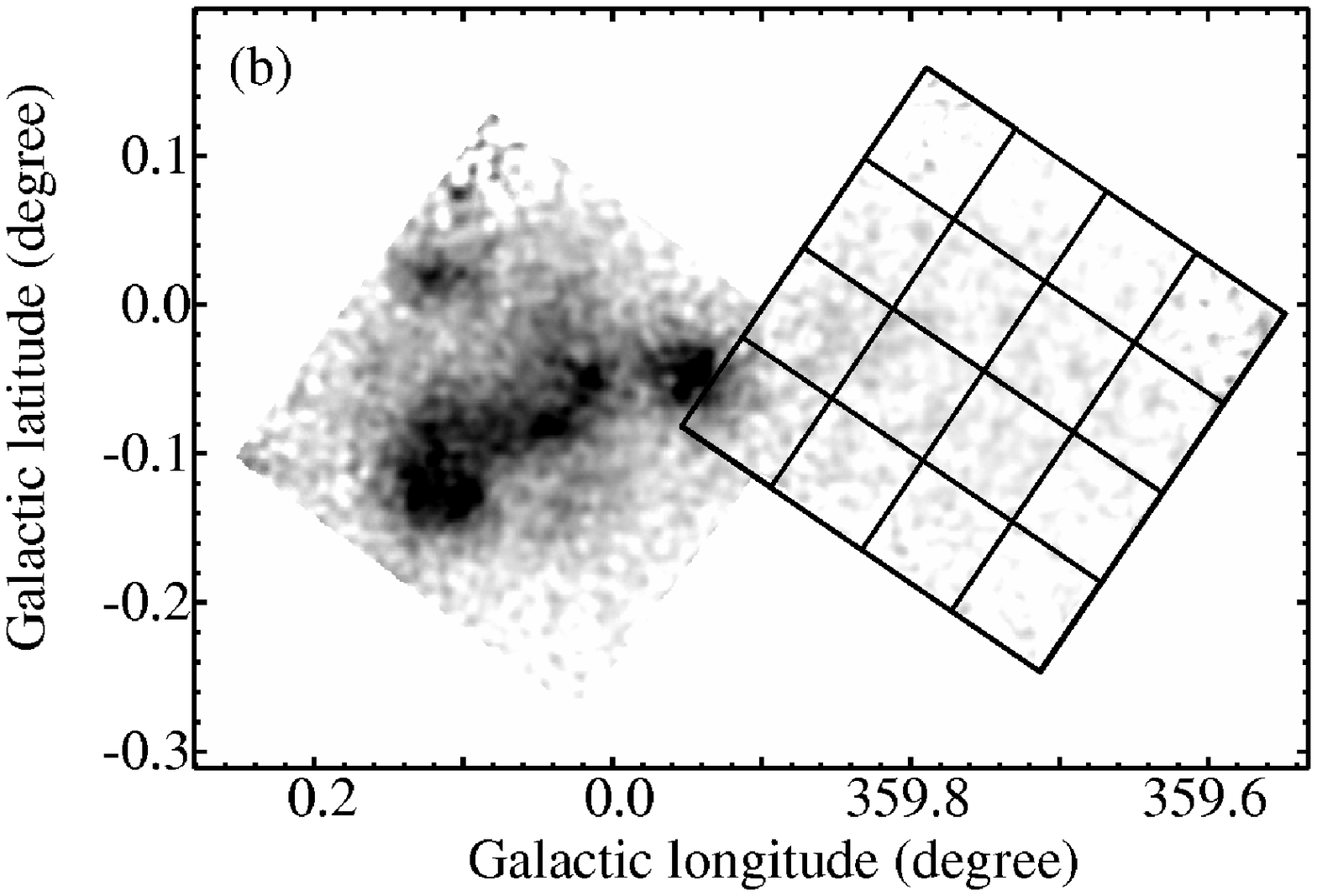}
  \end{center}
\caption{
Mosaic (2-pointings of the east and west fields) maps of the
6.7~keV line (a) and of the 6.4~keV line (b) bands. The east
and west fields are the left (east), and the right (west)
sides, respectively. Solid polygons in figure (a) are
6.4~keV clumps (here, the west clump is source 1, and the
east clump is source 2 ), while the dashed ellipse is the
background region. The background data for the 6.4~keV
clumps were obtained by excluding the data of source 1 and
2. The grid in the west field shows 16 segmentations for the
spatial resolved study (see text).}

\label{fig:IMAGE}
\end{figure}

\section{Analyses and Results}
\subsection{The Extended Emission near the GC (GCDX)}

\begin{figure*}[p] 
  \begin{center}
    \FigureFile(70mm,50mm){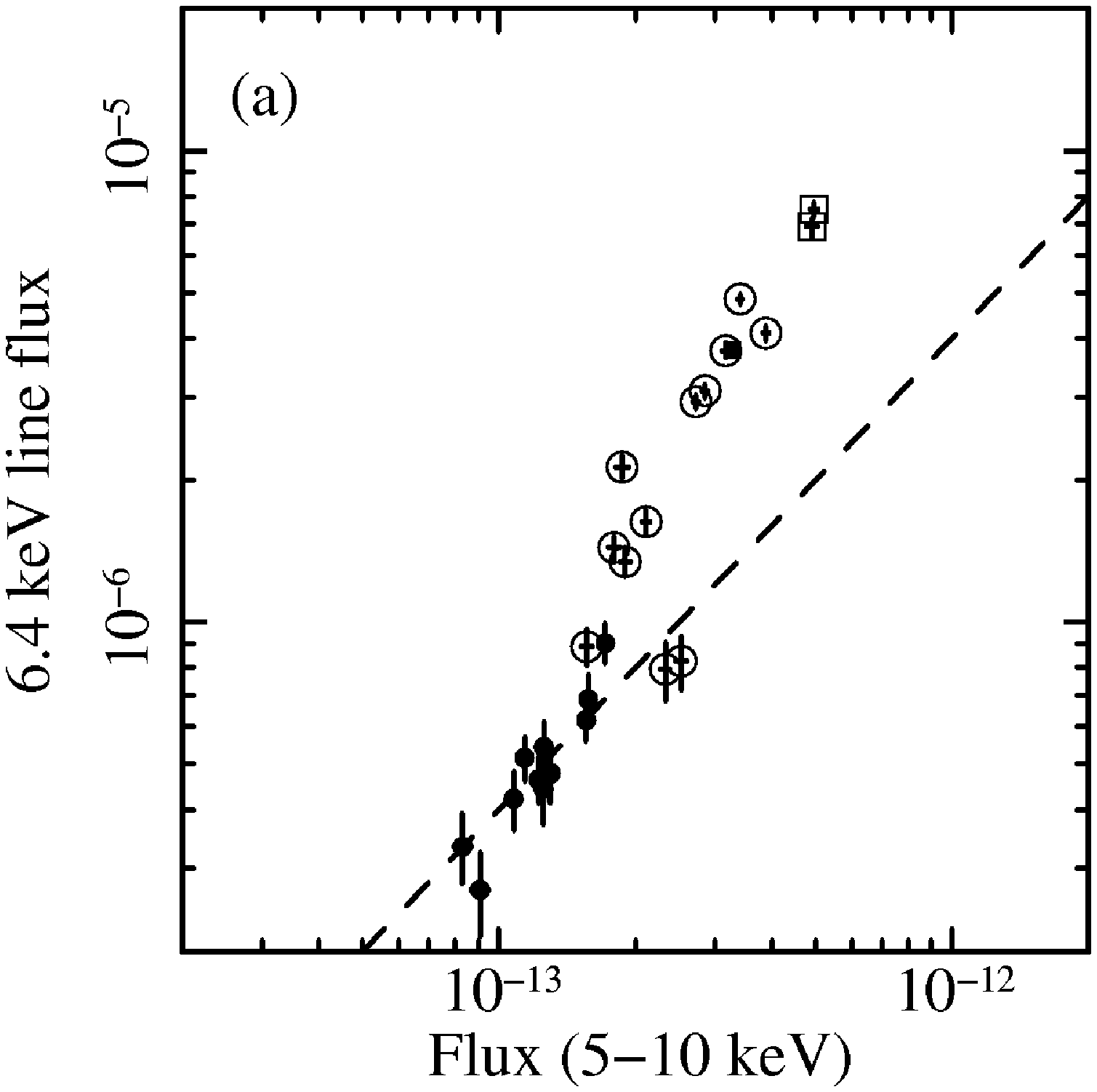}
    \FigureFile(70mm,50mm){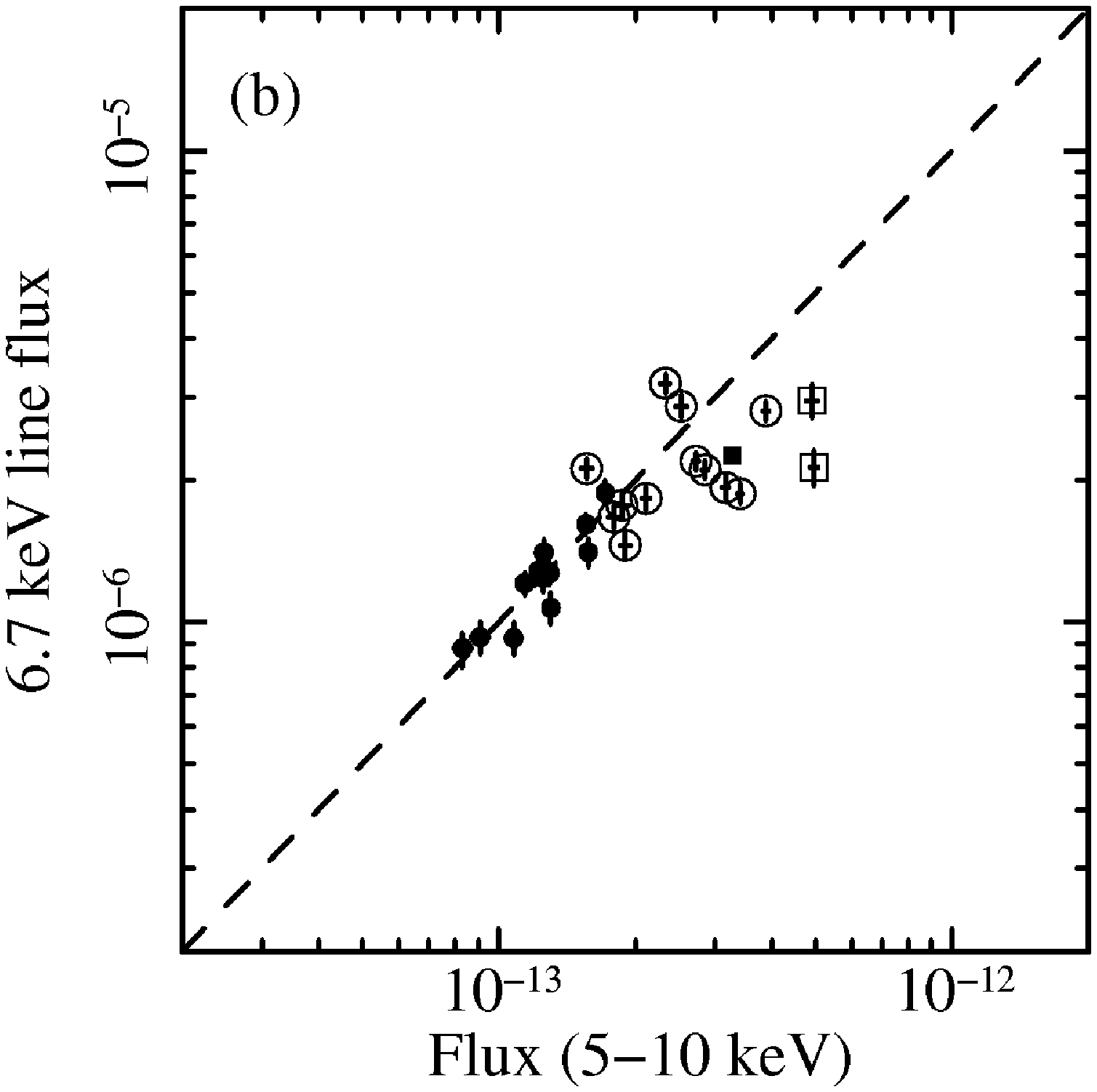}
    \FigureFile(70mm,50mm){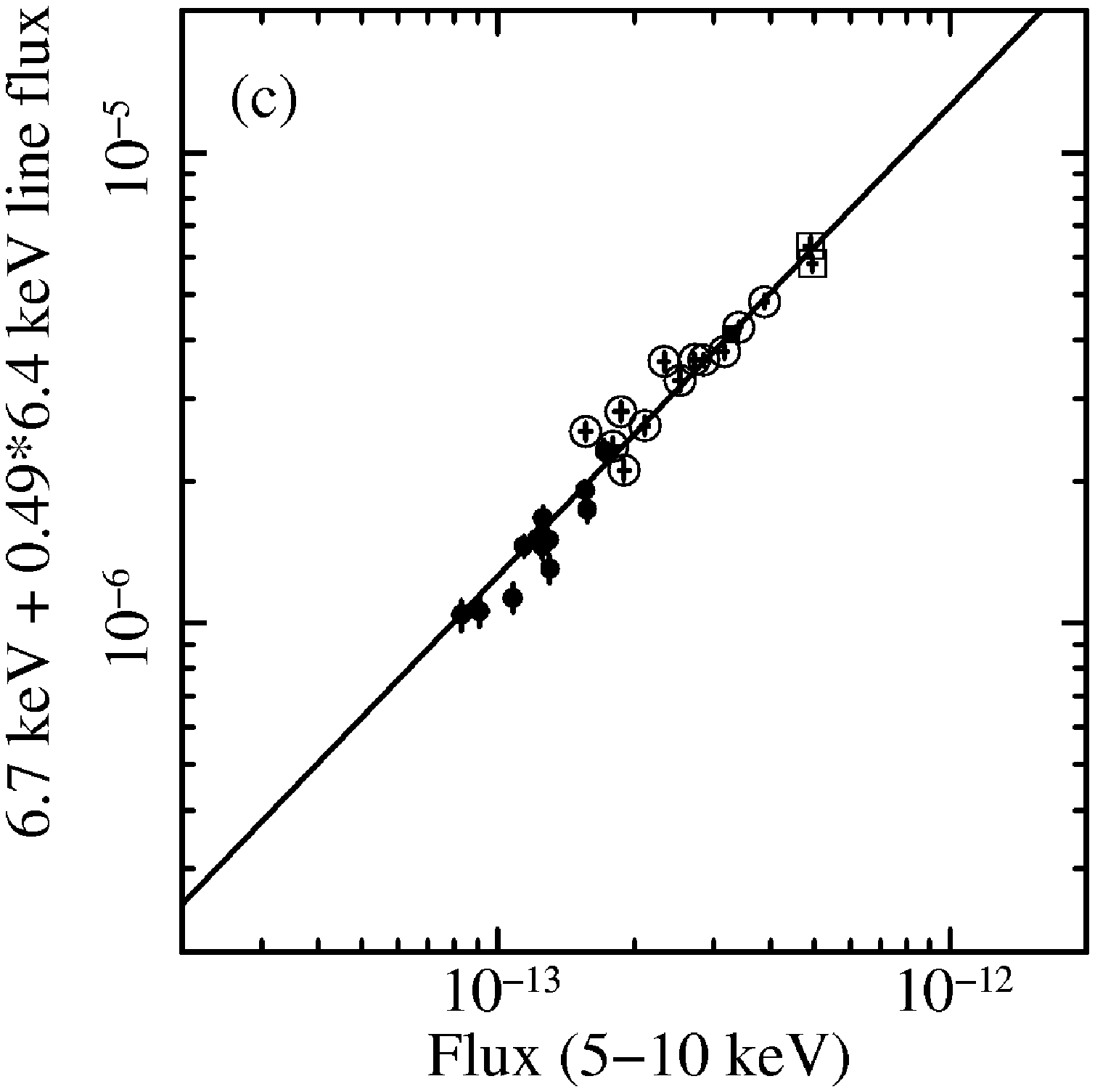}
    \FigureFile(60mm,50mm){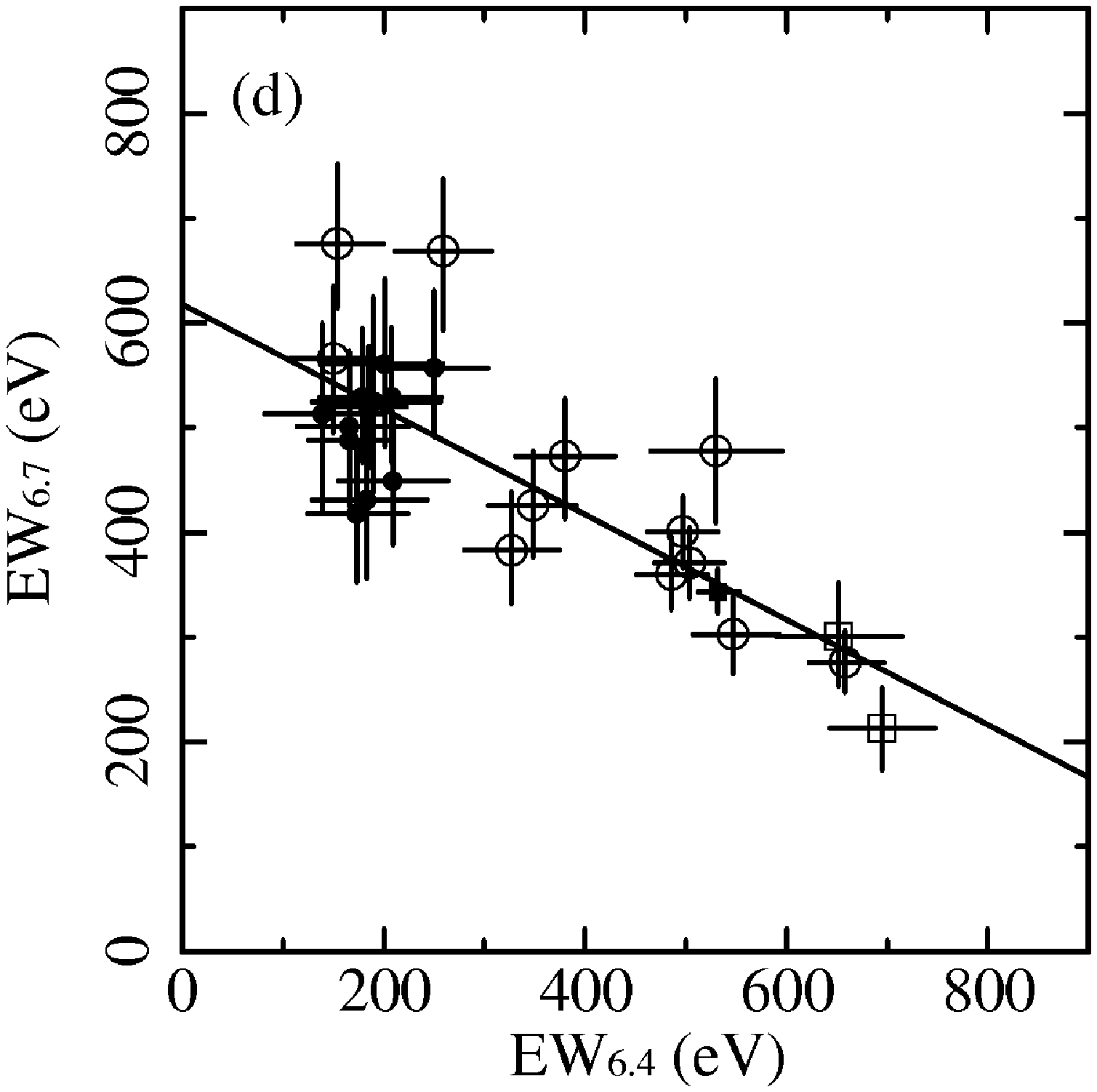}
\end{center}
\caption{ 

Correlation plots between the physical parameters. The open
and filled circles are data from the east and west fields,
respectively. For comparisons, the data of the 6.4~keV
clumps (sources 1 and 2) and background are also plotted, as
shown by the open and filled squares, respectively (see
section 3.2).\\

a) Plots of the 5--10~keV band flux ($L_{5-10}$, horizontal
axis) vs. the Fe\emissiontype{I}-K$\alpha$ flux ($F_{6.4}$,
vertical axis). The unit of $F_{6.4}$ and $L_{5-10}$ are
[photons cm$^{-2}$ s$^{-1}$ arcmin$^{-2}$] and
[ergs~cm$^{-2}$ s$^{-1}$ arcmin$^{-2}$], respectively.  The
dashed line is an eye guide to the proportional relation of
$L_{5-10}$~$\propto$~$F_{6.4}$.\\

b) Same as (a), but for Fe\emissiontype{XXV}-K$\alpha$
($F_{6.7}$).\\

c) Same as (a), but for the sum of the line flux of
Fe\emissiontype{XXV}-K$\alpha$ ($F_{6.7}$) and
Fe\emissiontype{I}-K$\alpha$ ($F_{6.4}$). The solid line is
the best-fit proportional line of
$F_{6.7}$+0.49$\times$$F_{6.4}$
=1.26$\times10^{7}\times$$L_{5-10}$. \\ 

d) Same as (a), but for the relation of the equivalent width
of the Fe\emissiontype{XXV}-K$\alpha$ line ($EW_{6.7}$) and
that the Fe\emissiontype{I}-K$\alpha$ line ($EW_{6.4}$). The
solid line shows the best-fit relation of
$EW_{6.7}$+0.50$\times$$EW_{6.4}$ =0.62~~[keV].

}

\label{fig:cor}
\end{figure*}

In order to see the spatial distribution of the Fe K-shell
lines, we made line images of the 6.7~keV
(Fe\emissiontype{XXV}-K$\alpha$) and 6.4~keV
(Fe\emissiontype{I}-K$\alpha$) lines with the respective
energy bands of 6.62--6.78~keV and 6.32--6.48~keV.  The
images are shown in figure
\ref{fig:IMAGE}, where they include both the relevant
line flux and the underlying continuum flux.  From figure
\ref{fig:IMAGE}, we can see that the 6.7~keV line flux in the
east field is systematically larger than that in the west
field.  This contrast is clearer in the 6.4~keV line band,
showing clear clumps near $(l, b) = (0.03, -0.07)$ and
$(0.12, -0.12)$ (source 1 and 2; the polygons in figure 1a).

To study more quantitatively, we divided both the east and
west fields into 16 segments each (32 segments for total),
as given by the grid in figure~1b.  Since the 4 segments in
each field corner are partially contaminated by the
calibration X-rays (Mn\emissiontype {I}-K$\alpha$ and
K$\beta$ lines at 5.9 and 6.5 keV, respectively), we made
the X-ray spectra for the remaining 24 segments in the
5--11.5~keV band, and fitted with a phenomenological model
as follows:
\begin{eqnarray}
&{\rm Abs} \times ({\rm PL}+ {\rm Abs} \times {\rm CXB} + {\rm Gaussians}) \nonumber&\\
&& \hspace{-10em} \left[\rm{photons~keV}^{-1}~{\rm cm}^{-2}~{\rm s}^{-1}~{\rm str}^{-1}\right],
\end{eqnarray}
where PL is a power-law function, PL$=A \times
E^{-\mit\Gamma}$.  CXB is the cosmic X-ray background
modeled as CXB=$8.75\times (E/1~{\rm keV})^{-1.486}$.  Abs
is the intra-Galactic absorption in the line of sight to the
GC, and is given by $e^{-\sigma(E)N_{\rm H}}$, where $N_{\rm
H}$ and $\sigma(E)$ are, respectively, the hydrogen column
density and the absorption cross section with the solar
abundances. The GCDX suffers due to the Galactic absorption
(Abs) on the front side of the GC, while the CXB suffers due
to both (front and back) side absorptions; hence, Abs is
applied twice to the CXB, as is explicitly given in
equation~1. Gaussians are given as $(F/\sqrt{2 \pi
w})e^{-(E-E_{\rm C})^2/2w^2}$, where $w$ is the intrinsic
line width ($1\sigma$).

Following \citet{Ko07d}, we employed 10 K-shell lines (10
Gaussians) due to highly ionized and neutral atoms.  The
brightest 4 lines are K$\alpha$ from neutral
(Fe\emissiontype{I}), He-like (Fe\emissiontype{XXV}) and
H-like (Fe\emissiontype{XXVI}) irons and K$\beta$ of
Fe\emissiontype{I}, at 6.4, 6.7, 6.97 and 7.06~keV,
respectively. The other weak lines are K$\alpha$ from
neutral (Ni\emissiontype{I}) and He-like
(Ni\emissiontype{XXVII}) nickel, K$\beta$ and K$\gamma$ from
Fe\emissiontype{XXV} and Fe\emissiontype{XXVI}, at 7.47,
7.81, 7.88, 8.25, 8.29, and 8.70~keV, respectively.

At first, we fitted the spectra from the full region of the
east and west fields separately, with essentially the same
fitting procedures as those of \citet{Ko07d}: the
line center energy, width and normalization (flux) of
Fe\emissiontype{I}-K$\beta$ are fixed to 1.103, 1.103 and
0.11 times to those of Fe\emissiontype{I}-K$\alpha$ (see
\cite{Ko07d}). The widths ($w$) for the 6 weak lines were
fixed to be 1~eV (narrow line approximation). Then, using the
best-fit line widths ($w$) and the line center energies ($E_{\rm
C}$), we fitted the 24-segment spectra.  The line flux ratio
of Fe\emissiontype{I}-K$\beta$/Fe\emissiontype{I}-K$\alpha$
was fixed to the theoretical value of 0.11 (see
\cite{Ko07d}).  Therefore, the free parameters are normalizations
of the power-law component ($A$) and those of the emission
lines ($F$) except for Fe\emissiontype{I}-K$\beta$, $N_{\rm
H}$ and the power-law index ($\mit\Gamma$).

Using the best-fit photon indices ($\mit\Gamma$), the line
fluxes of 6.4~keV ($F_{6.4}$) and 6.7~keV ($F_{6.7}$), the
energy flux in the 5--10~keV band ($L_{5-10}$), and the
equivalent widths of the 6.4~keV ($EW_{6.4}$) and the 6.7
keV ($EW_{6.7}$) lines, we made correlation plots
(figure~\ref{fig:cor}). For a consistency check of the
best-fit parameters between the east and west fields, we
made two spectra from the small region of the overlap of
both fields and fitted the spectra with the same model of
equation 1.  Since this small region includes the
calibration Mn\emissiontype{I}-K$\alpha$ and K$\beta$ lines
at 5.9 and 6.5 keV, the 6.4 keV flux ($F_{6.4}$) is
contaminated by the 6.5 keV line.  Therefore, we compared
the best-fit flux of the 6.7 keV line ($F_{6.7}$) and the
power-law in the 5--10~keV band ($L_{5-10}$).  The best-fit
values of $F_{6.7}$ are $3.88_{-0.16}^{+0.40}$ and
$4.06_{-0.28}^{+0.21}$ (in unit of $10^{-5}$~photons
cm$^{-2}$ s$^{-1}$) for the east and west fields,
respectively. The best-fit values of $L_{5-10}$ are
$3.57_{-0.09}^{+0.09}$ and $3.58_{-0.09}^{+0.08}$ (in unit
of $10^{-12}$~ergs~cm$^{-2}$~s$^{-1}$) for the east and west
fields, respectively. Thus, we confirmed that the relevant
best-fit parameters obtained from two fields are consistent
with each other.

Figure~\ref{fig:cor}a shows that the flux ratio of the
Fe\emissiontype{I}-K$\alpha$ line ($F_{6.4}$) to the
5--10~keV band ($L_{5-10}$) is not constant, but $F_{6.4}$
shows excess at larger flux domains.  Figure~\ref{fig:cor}b
shows a vice versa flux-relation of the
Fe\emissiontype{XXV}-K$\alpha$ line ($F_{6.7}$) to the
5--10~keV band ($L_{5-10}$). These facts indicate that the
5--10~keV band flux ($L_{5-10}$) does not solely associate
with the Fe\emissiontype{I}-K$\alpha$ line nor
the Fe\emissiontype{XXV}-K$\alpha$ line.

We therefore searched for a possible combination of the
Fe\emissiontype{XXV}-K$\alpha$ and
Fe\emissiontype{I}-K$\alpha$ flux to become proportional to
the 5--10~keV band flux. Figure \ref{fig:cor}c shows the
relation of the combined 6.7 keV and 6.4 keV flux ($F_{6.7}$
and $F_{6.4}$) vs. the 5--10 keV band flux ($L_{5-10}$).  The
best-fit relation is
\begin{equation}
F_{6.7}+0.49(_{-0.04}^{+0.03}) \times F_{6.4} 
= 1.26 \times 10^{7} \times L_{5-10},
\end{equation}
where the data dispersion ($1 \sigma$) from the best-fit
relation is 11\%.  This relation is confirmed by 
correlation plots of the equivalent width of the
Fe\emissiontype{XXV}-K$\alpha$ line ($EW_{6.7}$) and that of
Fe\emissiontype{I}-K$\alpha$ ($EW_{6.4}$). The solid line in
figure~\ref{fig:cor}d is the best-fit relation, given as
\begin{equation}
EW_{6.7} + 0.50(\pm0.06) \times EW_{6.4} 
= 0.62 (\pm0.07)~[{\rm keV}].
\end{equation}
We note that \citet{War06} reported a similar (but rather
qualitative) analysis using the emission lines obtained by
XMM-Newton observations.

\begin{figure} 
  \begin{center}
\FigureFile(80mm,50mm){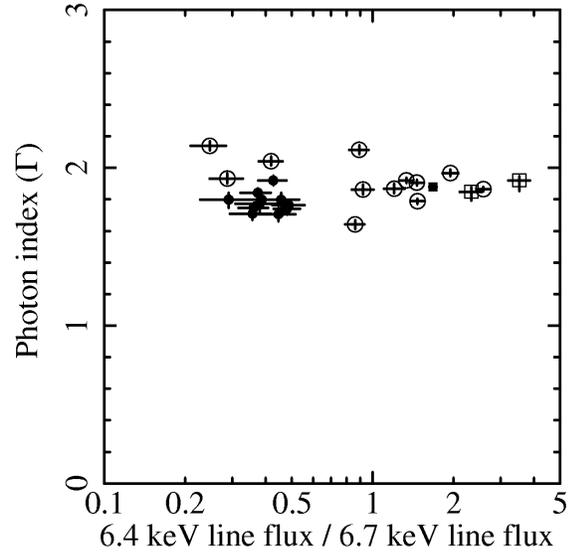}
\end{center}
\caption{Same as figure 2, but for the line flux ratio of 
Fe\emissiontype{I}-K$\alpha$ and 
Fe\emissiontype{XXV}-K$\alpha$ ($F_{6.4}/F_{6.7}$)  (horizontal axis) vs. the photon index $\mit\Gamma$ (vertical axis). 
}
\label{fig:gamma}
\end{figure}

The correlations of equations 2 and 3 suggest that the
power-law component (PL) can be divided into two parts, PL1
and PL2, which are the power-law continuums associated with
the K-shell lines from neutral and highly ionized atoms,
respectively. We hence divide the 5--10~keV band flux of
$L_{5-10}$ to $L1_{5-10}$ and $L2_{5-10}$, which belong to
PL1 and PL2, respectively.  These phenomenological relations
of equations 2 and 3 also mean that the flux ratio
$L2_{5-10}$/$L1_{5-10}$ is proportional to
$\sim$(1/0.5)$\times$ ($F_{6.7}$/$F_{6.4}$).  In figure
\ref{fig:gamma}, we plot the photon index ($\mit\Gamma$) as
a function of the flux ratio of Fe\emissiontype{I}-K$\alpha$
to Fe\emissiontype{XXV}-K$\alpha$
($F_{6.4}/F_{6.7}$). Although the values $F_{6.4}/F_{6.7}$
scatter largely from $\sim$0.2 to 4, $\mit\Gamma$ is almost
constant at about 1.9; the continuum shape is the same
regardless the line ratio. Therefore, PL1 and PL2 have
nearly the same photon indices ($\mit\Gamma$) of 1.9.

\subsection{X-ray  Image and Spectra of the 6.4~keV Clumps near the GC} 

In figure \ref{fig:IMAGE}, we can see strong enhancements of
the 6.4~keV line near at the Radio Arc (sources 1 and 2). To
make reliable spectra of sources 1 and 2, a precise
estimation of the Galactic center diffuse X-rays (GCDX) is
particularly important, because the GCDX comprises the major
background, and is variable from position to position (see
section 3.1). To minimize any systematic error due to
subtraction of the position-dependent GCDX, we selected the
background region to be as near as possible to sources 1 and
2.  The background region thus selected is shown by the
dashed ellipse in figure \ref{fig:IMAGE}, where the data of
sources 1 and 2 (polygons) are excluded.

First, we obtained the source and background spectra, and
fitted with a phenomenological model of equation 1. The
best-fit fluxes of the Fe\emissiontype{XXV}-K$\alpha$ lines
for sources 1 and 2, and that of the background are
2.95($_{-0.24}^{+0.24})\times 10^{-6}$,
2.13($_{-0.18}^{+0.19})\times 10^{-6}$ and
2.25($_{-0.06}^{+0.06})\times 10^{-6}$
[photons~cm$^{-2}$~s$^{-1}$~arcmin$^{-2}$], respectively.
Therefore, the GCDX in source 1 would be larger, but that in
source 2 is smaller than the GCDX in the background region.
Thus, a key issue is how to properly subtract the GCDX. As
is suggested in section 3.1, the power-law component (PL) of
the GCDX is divided into PL1 and PL2, which are associated
with the 6.4 keV (neutral iron) and 6.7 keV (He-like iron)
lines, respectively.  It is very likely that PL1 and PL2 are
also associated with the K-shell lines from neutral and
highly ionized atoms.  In order to subtract PL2 and
associated K-shell lines from highly ionized atoms, we
introduced a multiply factor, $\alpha$, which is the ratio
of the 6.7~keV flux ($F_{6.7}$) of source 1 (or 2) to that
of the background spectrum.  The factors $\alpha$ are 1.31
for source 1 and 0.95 for source 2.  We then re-constructed
a background model consisting of the K-shell lines from
highly ionized atoms and the relevant continuum
component. As for the fluxes of the K-shell lines from
highly ionized atoms, we multiplied the factor $\alpha$ to
those of the best-fit line-flux of the background spectrum.
For the continuum, on the other hand, we multiplied the
factor of $\alpha \times F_{6.7}/(F_{6.7}+0.5 \times
F_{6.4})$ to the best-fit PL of the background spectrum,
where $F_{6.7}$ and $F_{6.4}$ are the line fluxes of the
background regions. This is the same as $\alpha \times {\rm
PL2}$, where PL2 is that from the background region.

\begin{figure} 
  \begin{center}
    \FigureFile(80mm,50mm){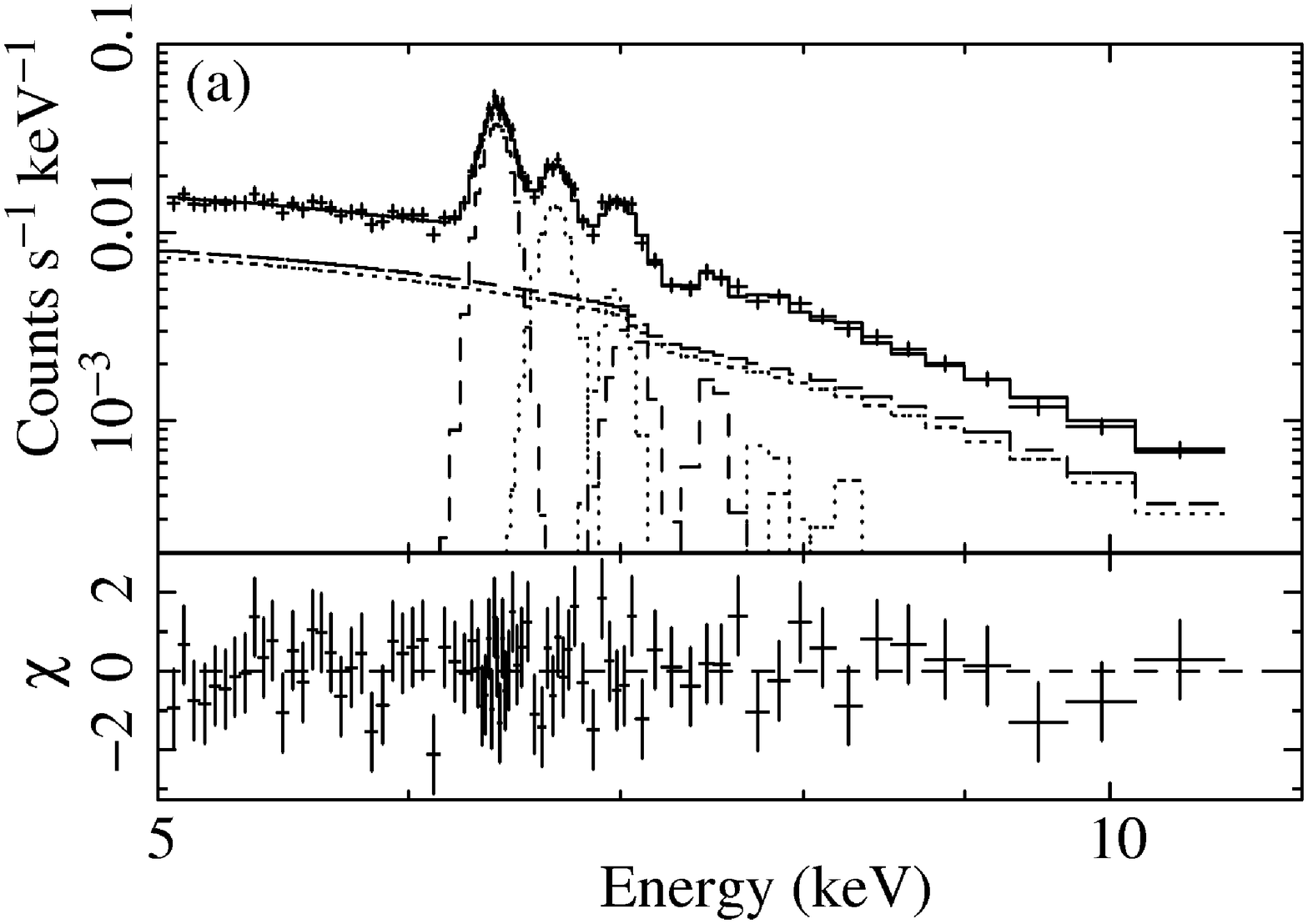}
    \FigureFile(80mm,50mm){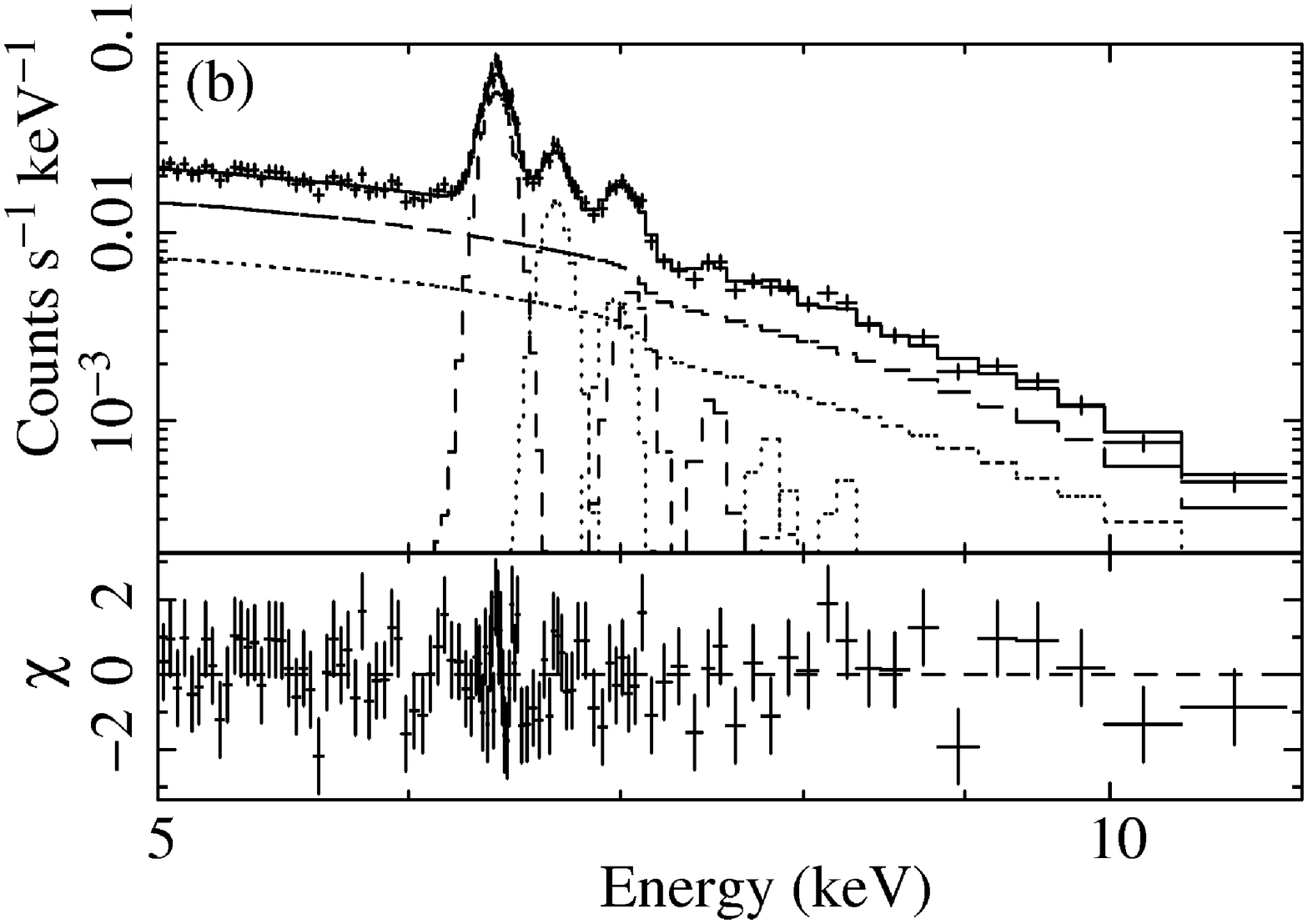}
  \end{center}
\caption{
X-ray spectra of source 1 (a) and source 2 (b). The
dashed lines are the best-fit model of the sources, while
the dotted lines are the model GCDX spectra obtained by the
method given in the text. The CXB spectra are out of the
frame of these figures.
\label{fig:SS}}
\end{figure}

Adding this model background and CXB, we fitted the spectra
of sources 1 and 2 with a model of an absorbed power-law
plus K$\alpha$ and K$\beta$ lines of neutral iron and
nickel.  The best-fit spectra are given in figure
\ref{fig:SS} with the dashed lines together with the model
background (dotted line). The best-fit source parameters are
listed in table~1. Note that the spectral parameters of
sources 1 and 2 include the PL1 components of the background
region.

\begin{table} 
  \begin{center}
    \caption{Best-fit parameters of the background-subtracted sources.}
    \label{tab:SS}
    \begin{tabular}{lcc}
      \hline          
      & source 1	& source 2  \\
      \hline                             
$Ab_{\rm Fe}$\footnotemark[$*$]
&3.8$_{-0.5}^{+0.6}$  &3.9$_{-0.6}^{+0.4}$   \\
Photon index ($\mit\Gamma$) 
        &1.83$_{-0.03}^{+0.03}$&1.86$_{-0.02}^{+0.03}$ \\
$L_{5-10}$\footnotemark[$\dagger$]
           &2.61$_{-0.16}^{+0.14}$&3.37$_{-0.12}^{+0.12}$ \\
$F_{6.4}$\footnotemark[$\ddagger$] 
&6.92$_{-0.30}^{+0.29}$&7.45$_{-0.23}^{+0.29}$ \\
$EW_{6.4}$\footnotemark[$\S$] 
&1.23$_{-0.14}^{+0.14}$ &1.03$_{-0.09}^{+0.08}$ \\
$F_{7.05}$\footnotemark[$\ddagger$]   
&0.99$_{-0.25}^{+0.27}$&1.02$_{-0.23}^{+0.19}$ \\
      \hline
      \multicolumn{3}{@{}l@{}}{\hbox to 0pt{\parbox{70mm}{\footnotesize
	    \par\noindent
	    \footnotemark[$*$] Iron abundances determined by the iron K-edge depth
with a fixed $N_{\rm H}$ of 6$\times$10$^{22}$cm$^{-2}$. 
	    \par\noindent
	    \footnotemark[$\dagger$] Unabsorbed 5--10~keV band flux in unit of 
10$^{-13}$~erg~s$^{-1}$~cm$^{-2}$ arcmin$^{-2}$.
           \par\noindent
	    \footnotemark[$\ddagger$] Unabsorbed line fluxes of
Fe\emissiontype{I}-K$\alpha$ ($F_{6.4}$) and K$\beta$ ($F_{7.05}$) 
in unit of 10$^{-6}$photons s$^{-1}$~cm$^{-2}$ arcmin$^{-2}$.    
           \par\noindent
	    \footnotemark[$\S$] Equivalent width
of the Fe\emissiontype{I}-K$\alpha$ line in unit of~keV.}\hss}}
\end{tabular}
\end{center}
\end{table}

\subsection{Timing Analysis of the 6.4~keV Clumps near the GC}

\begin{figure} 
  \begin{center}
    \FigureFile(80mm,50mm){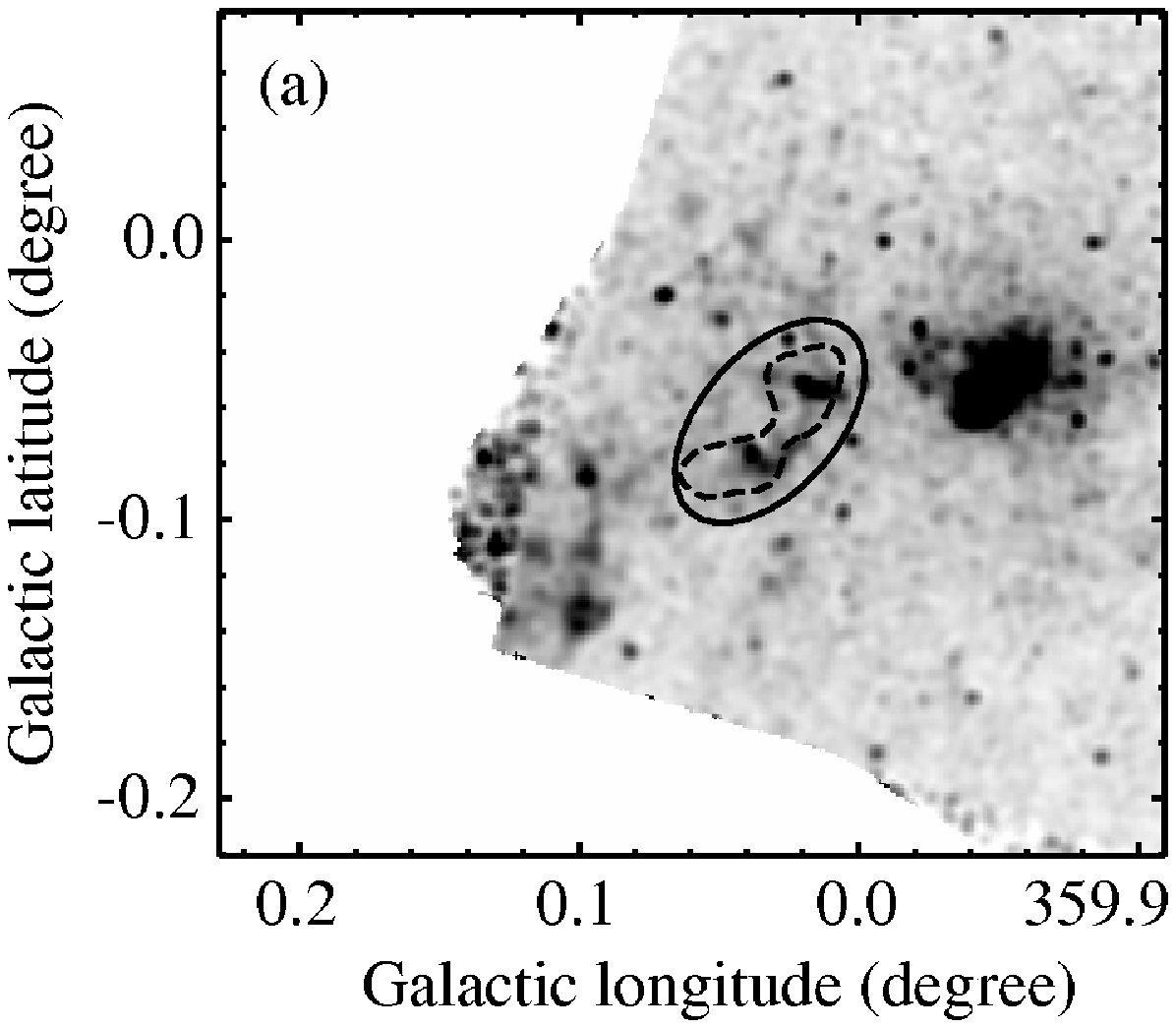}
    \FigureFile(80mm,50mm){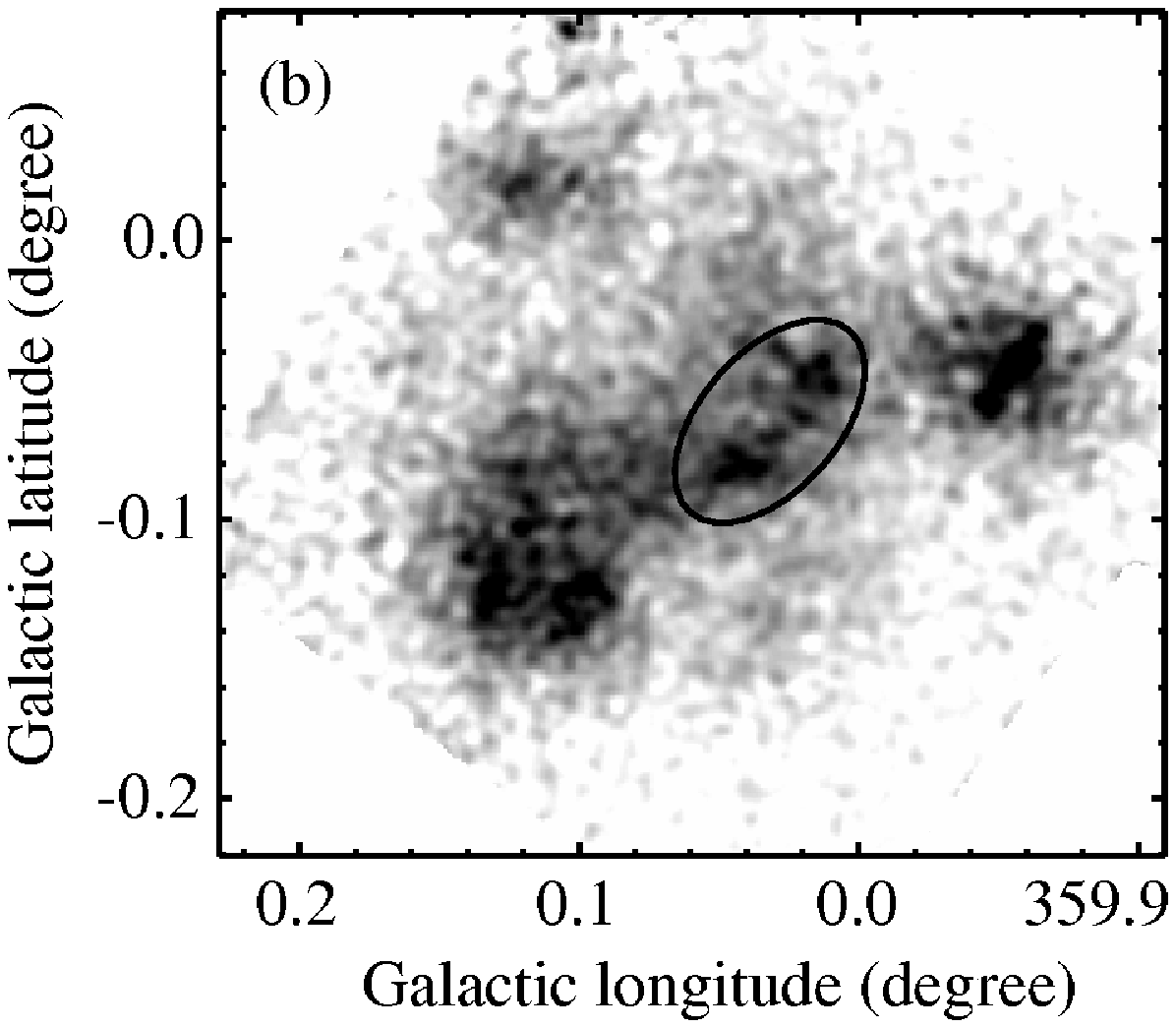}
  \end{center}
  \caption{(a) X-ray image obtained with Chandra in the 6--7~keV band, 
    and (b) the Suzaku image in the 6.4~keV line band (6.32-- 6.48 keV). The dashed polygon in (a) shows 
    source 1 (see text). The spectra are taken from the solid ellipses in (a) and (b).}
  \label{fig:S_C}
\end{figure}

\citet{Mu07} reported a time variability of the sub-structures 
in source 1 in Chandra observations. We therefore extended
the time variability study for a longer time scale from the
Chandra (2002) to the Suzaku (2005) observations. Since the
spatial resolution of Suzaku is limited to resolve the
sub-structures in source 1, and since the image position of
source 1 with Suzaku and Chandra are slightly shifted from
each other, we extracted the X-ray spectra from a larger
region than source 1, as is given by the solid ellipses in
figure~\ref{fig:S_C}.  We subtracted the NXBG from the
Suzaku spectrum in the same way as previously described, and
fitted the spectra with the model of equation 1.  The
best-fit Suzaku fluxes of the 6.4 and 6.7~keV lines are
given in table~\ref{tab:S_C}. For the Chandra spectrum, we
subtracted the off-plane CXB (including NXBG) data using the
"blank-sky" data-sets. The Chandra spectrum was fitted with
the same model as Suzaku (equation 1), fixing the line
energies, the power-law index and iron K-edge absorption to
the Suzaku best-fit values. Free parameters were
normalizations (flux) of the power-law and Gaussian lines.
For a reasonable fit, we fine-tuned the Chandra energy
gain by $\sim$0.2\%.  The best-fit Chandra fluxes of the 6.4
and 6.7~keV lines are listed in table~\ref{tab:S_C}.

From table~\ref{tab:S_C}, we can see that the 6.7~keV line
flux is constant within the 90\% level of the statistical
errors. This result is reasonable, because the 6.7 keV line
is due to the largely extended GCDX, and hence should be
invariant on the time scale of a few years.  In other words,
the constant 6.7~keV flux supports that the over-all
systematic flux error of the 6.4~keV and 6.7~keV lines
between the Chandra and the Suzaku observations, under the
present procedure of data selection, screening and analysis,
is smaller than the statistical 1.5$\sigma$ error. We note
that even if the power-law index and iron K-edge absorption
are free parameters in the fitting of the Chandra spectrum,
the best-fit fluxes have almost the same values as listed in
table \ref{tab:S_C}.  Thus, from table
\ref{tab:S_C}, the flux change of the 6.4~keV line from the
Chandra (2002) to the Suzaku (2005) observations is
significant at the 4.7$\sigma$ level (note that the errors
in table 2 are at the 90\% level).

\begin{table} 
  \begin{center}
    \caption{Best-fit fluxes of the 6.4 and 6.7~keV lines of the Suzaku and Chandra observations.}
    \label{tab:S_C}
    \begin{tabular}{lcc}
      \hline          
       &  Chandra (2002)&  Suzaku  (2005)\\
       \hline
      $Ab_{\rm Fe}$\footnotemark[$*$]
 &3.4\footnotemark[$\dagger$]& 3.4$_{-0.4}^{+0.5}$\\
      Photon index ($\mit\Gamma$)
  &1.77\footnotemark[$\dagger$]& 1.77$_{-0.02}^{+0.02}$ \\
$F_{6.4}$\footnotemark[$\ddagger$]
&7.83$_{-0.23}^{+0.23}$&6.89$_{-0.23}^{+0.20}$\\
$F_{6.7}$\footnotemark[$\ddagger$]
&3.37$_{-0.20}^{+0.20}$ &3.61$_{-0.18}^{+0.19}$\\
       \hline
      \multicolumn{3}{@{}l@{}}{\hbox to 0pt{\parbox{80mm}{\footnotesize
         \par\noindent
            \footnotemark[$*$] Iron abundances determined by the iron K-edge depth
with a fixed $N_{\rm H}$ of 6$\times$10$^{22}$cm$^{-2}$.
  	    \par\noindent
	    \footnotemark[$\dagger$] Fe abundance and $\mit\Gamma$ are fixed to the Suzaku best fit values.
	    \par\noindent
	    \footnotemark[$\ddagger$] Unabsorbed line fluxes of  Fe\emissiontype{I}-K$\alpha$  and
Fe\emissiontype{XXV}-K$\alpha$ in unit of 10$^{-5}$photons 
s$^{-1}$cm$^{-2}$.}\hss}}
    \end{tabular}
  \end{center}
\end{table}

\section {Discussion} 

\subsection{Decomposition of the GC Emission} 

The GCDX would have at least 3 components: high-temperature
plasma (component-1), the 6.4 line with Thomson scattering
or bremsstrahlung continuum (component-2), and integration
of point sources plus other possible origins (component-3).
K-shell lines from highly ionized atoms from component-1
constrain the plasma parameters. The most important lines
are Fe\emissiontype{XXV}-K$\alpha$ (the 6.7 keV line),
Fe\emissiontype{XXVI}-K$\alpha$ (the 6.97 keV line) and
Fe\emissiontype{XXV}-K$\beta$ (the 7.88 keV line).  The flux
ratio of the 6.97~keV line to the 6.7~keV line gives the
ionization temperature, while that of the 7.88~keV line to
the 6.7~keV line gives the electron
temperature. \citet{Ko07d} extensively studied the line flux
(and ratio) from the GC (the east and west fields), and
concluded that the GCDX has a high-temperature plasma in
collisional ionization equilibrium.  The GC spectrum was, in
fact, nicely fitted with a 6.5 keV plasma and a power-law
with a photon index of ${\mit\Gamma}=1.4$, plus neutral
K-shell lines (see figure 7 in \cite{Ko07d}). The continuum
flux in the 5--10 keV band of the high-temperature plasma
was $\sim$0.5$\times$$L_{5-10}$ of the GCDX (the east and
west fields) and that of the power-law ($\mit\Gamma$=1.4)
was $\sim$0.5$\times$$L_{5-10}$ (see figure 7 in
\cite{Ko07d}).

As we already proposed, we can decompose the 5--10~keV band
flux ($L_{5-10}$) into two components, $L1_{5-10}$ and
$L2_{5-10}$ with the flux ratio $L2_{5-10}$/$L1_{5-10}$
being proportional to (1/0.5)$\times$ ($F_{6.4}$/$F_{6.7}$).
\citet{Ko07d} found that the mean fluxes of the 6.7~keV line
($F_{6.7}$) and that of the 6.4 keV line ($F_{6.4}$) in the
east and west fields are nearly equal to each other (see
table~4 and figure~7 in \cite{Ko07d}).  Therefore, for the
mean 5--10~keV band flux of the east and west fields, about
2/3 (1/1.5) is attributable to the 6.7~keV line and the
other 1/3 (0.5/1.5) is to the 6.4~keV line. The photon
indices ($\mit\Gamma$) for the both components are the same
at 1.9 (see section 3.1).
 
\begin{figure} 
  \begin{center}
    \FigureFile(80mm,50mm){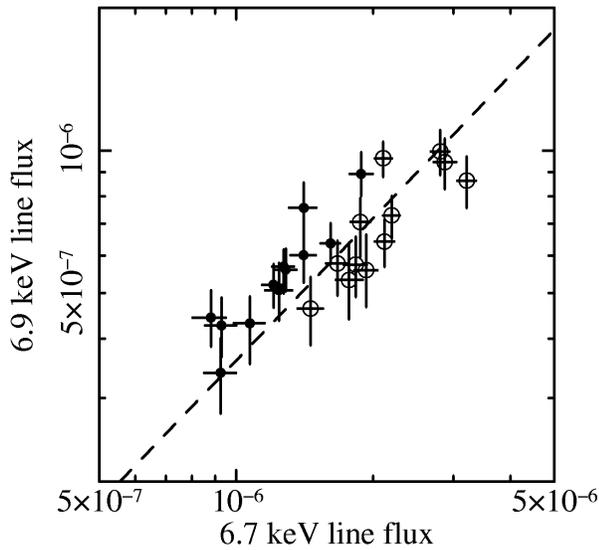}
  \end{center}
  \caption{ Same as figure 2, but the line flux plots of
Fe\emissiontype{XXVI}-K$\alpha$ ($F_{6.97}$)  (horizontal axis) 
and Fe\emissiontype{XXV}-K$\alpha$ ($F_{6.7}$) (vertical axis).   \\
}
\label{fig:Temp}
\end{figure}

The above two decompositions implicitly assumed that the
6.7~keV line is due to a plasma with a spatially uniform
temperature. This assumption was verified by a spatial
analysis of the flux ratio of Fe\emissiontype{XXV}-K$\alpha$
(6.7~keV line, $F_{6.7}$) to Fe\emissiontype{XXVI}-K$\alpha$
(6.97~keV line, $F_{6.97}$).  In figure \ref{fig:Temp}, we
plot the correlation of $F_{6.97}$ and $F_{6.7}$.

From figure~\ref{fig:Temp}, we conclude that the plasma
temperatures are approximately uniform. In detail, however,
the flux ratios are systematically larger, and hence show a
higher temperature in the west field than that in the east
field by $\sim$10\% (figure \ref{fig:Temp}).
\citet{Ko07d} reported that the GC has a lower temperature 
plasma (the soft component) with highly ionized atomic
lines, such as silicon and sulfur.  We therefore fit the
low-energy band spectra, and found the plasma temperature to
be $\sim$1~keV.  The contribution of this plasma to the
6.7~keV-line flux is 5--10\%, but those to the 6.96~keV line
and the 5--10~keV flux are negligible. The flux shift of the
6.7~keV line is within the 1$\sigma$ dispersion of the line
flux vs. the continuum flux correlation
(figure~\ref{fig:cor}c). Including these effects increases
the temperature by $\sim$5\%.  Using the K-edge structure,
Koyama et al. (2007d) determined that the line-of-sight
$N_{\rm Fe}$ to the GCDX is
9.7$\times10^{18}$~cm$^{-2}$. This corresponds to a 3.5
solar abundance of iron, assuming that the $N_{\rm H}$ to
the GCDX is 6$\times10^{22}$~cm$^{-2}$.  However the assumed
$N_{\rm H}$ of 6$\times10^{22}$~cm$^{-2}$ may be smaller
than the typical values. In fact, the Suzaku observations on
the Sgr A East (SNR) and Arches (star cluster) revealed that
$N_{\rm H}$ is 9--14$\times10^{22}$ \citep{Ko07c,Ts07}.
Then, the iron K-edge abundance determined by the K-edge
absorption is reduced to be 2.3--1.5 solar. On the other
hand, the iron abundances of the 6.5 keV plasma in the GCDX
were determined to be $\sim$1 solar by \citet{Ko07d}.
\citet{War06} also reported that the iron abundance in the
GCDX plasma is one solar.  The 1~keV plasma is associated
with the 2.46~keV line. \citet{No08} and \citet{Mo08}
analyzed some of the 2.46~keV clumps, and found that the
abundances of iron and other heavy elements are consistent
with solar.  Thus, iron abundance in the 1~keV plasma is
likely to be one solar, and hence the $F_{6.7}$ values and
the temperatures may not be significantly changed.  We can
therefore ignore the 1~keV plasma in the discussion.

Now, we schematically show the results of the two different
decompositions of the 5--10~keV band flux ($L_{5-10}$) in
figure 7.  The right side shows nearly equal participation
of the 6.5~keV plasma (parenthesis is a phenomenological
photon index), and the power-law with ${\mit\Gamma}=1.4$
\citep{Ko07d}. The 6.4 keV line was treated separately, and
hence is not included in figure 7 (right). The left side
shows the phenomenological participations in this work, PL1
and PL2 with a flux ratio of 1:2.  The 6.7~keV and 6.4~keV
lines are mainly included in the white and grey regions,
respectively.  There is an apparent discrepancy between the
two decompositions.  This may be solved if we take a
point-source contribution into account (see the next
section).

\begin{figure} 
  \begin{center}
    \FigureFile(100mm,60mm){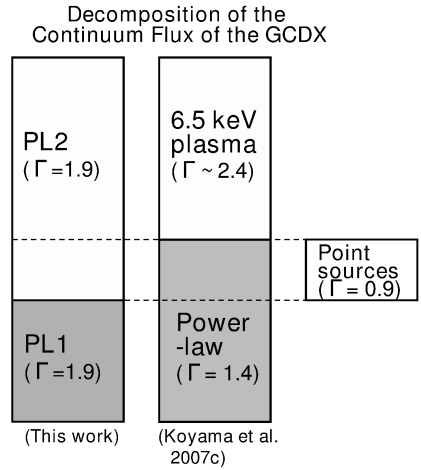}
  \end{center}
\caption{
Schematic picture of the continuum flux ($L_{5-10}$)
participation of the GCDX. The right side shows nearly equal
participation of the 6.5~keV plasma (parenthesis is a
phenomenological photon index), and the power-law
${\mit\Gamma}=1.4$ (\cite{Ko07d}).  The left side shows the
phenomenological participations (PL1 and PL2 with a ratio of
1:2).  White and grey indicates the region of the power-law,
which exhibits mainly the 6.7~keV and 6.4~keV lines,
respectively.
\label{fig:koyama}
}
\end{figure}

\subsection{Origin of the High Temperature Plasma} 

The spectral analysis of the GC (the east and west fields)
indicates that about half of the GCDX is due to the 6.5~keV
plasma, which emits the 6.7~keV line and other K-shell lines
from highly ionized iron and nickel (\cite{Ko07d}). However,
figure~\ref{fig:koyama} suggests that, additionally, at
least 1/6 of the total GCDX should also be the 6.7~keV line
emitter.  We propose that this is integrated point sources
(component-3), because \citet{Mu04} reported that similar
fractions of the GCDX come from point sources of $\ge3\times
10^{-15}$~ergs~cm$^{-2}$~s$^{-1}$ (2--9~keV), and the
spectrum has a strong 6.7~keV line and a rather weak 6.4~keV
line (see figure 7 of \cite{Mu04}).  Therefore, in the
present analysis, PL2 (the left of figure 7) may be
contaminated by the integrated flux of point sources. On the
other hand, in the analysis of \citet{Ko07d}, the
6.7~keV line from the point sources would be implicitly
included in the 6.5~keV plasma (the right), while the
continuum component of the point sources is included in the
power-law component of ${\mit\Gamma}=1.4$.  The continuum
shape of the 6.5~keV plasma is approximated by a power-low
of ${\mit\Gamma}=2.4$, while the integrated point sources
has 0.9 (\cite{Mu04}). Then, the flux-weighted mean value is
${\mit\Gamma} \sim 2.0$, consistent with ${\mit\Gamma}=1.9$ of
PL2. The power-law of ${\mit\Gamma}=1.4$ component is regarded
as a sum of the PL1 (${\mit\Gamma}=1.9$) and the point sources
of ${\mit\Gamma}=0.9$.  Then, the weighted mean photon index of
the power-law component becomes ${\mit\Gamma} \sim 1.6$,
consistent with ${\mit\Gamma}=1.4$.

From the above analysis, we infer that $\sim$1/6 of the GCDX
is due to point sources that are already resolved by Chandra
down to the flux
$3\times10^{-15}$~erg~cm$^{-2}$~s$^{-1}$. Since point
sources fainter than this flux level must be prevailing in
the GC, the fraction of point source contribution to the
GCDX of 1/6 is the lower limit.
\citet{Re06}, \citet{Re07a} and \citet{Re07b} 
proposed that the point-source population contributes
largely to the Galactic ridge diffuse X-rays (GRDX). This
scenario may also be applied to the GCDX. The point-source
distribution should be symmetric with respect to Sgr A*.
However, we found the east-west asymmetry of the 6.7~keV
line flux ($F_{6.7}$) as demonstrated by the open (east) and
filled (west) circles in figure 2b (also see figure 6 of
\cite{Ko07d}). The mean flux in the east field is
$\sim$1.5 times larger than that of the west field.  As we
already pointed out, the 1-keV temperature plasma can not
account for such a large asymmetry (see section 4.1). Thus,
the maximum possible contribution of the point sources to
the GCDX is about 70\% for the east region and 100\% for the
west field.  In this case, however, the spectrum of the
integrated point sources must have, approximately, a
power-law with $\mit\Gamma$=1.9 (or 17~keV temperature
plasma) with sizeable 6.4~keV and 6.7~keV lines ($EW_{6.4}$=
150--700~eV, $EW_{6.7}$ = 200--800~eV) and the line ratio
$F_{6.97}$/$F_{6.7}$=0.3--0.4.  As far as we know, the most
probable and popular point sources having such a hard
spectrum with sizeable 6.4, 6.7 and 6.96~keV lines are
intermediate poplars (IP). \citet{Ez99} complied 14 ASCA
data-sets of 12 IPs. The mean $EW_{6.4}$ and $EW_{6.7}$ are
140~eV and 220~eV, respectively (calculated from table 2 of
\cite{Ez99}), which are systematically smaller than those in
the GCDX (see figure 2d).  Thus, it may be unlikely that a
large fraction of the GCDX can be accounted for by such
point sources.  At this moment, however, we reserve any
definite conclusion until more quantitative estimates and
observations for both the point source and the diffuse
emission are evaluated.

\subsection{Origin of the 6.4~keV-line and the Clumps}

The origin of the 6.4~keV emission is due to the inner-shell
ionization of nearly neutral irons.  A plausible source for
the inner-shell ionization is bombarding on the cloud gas by
either high-energy electrons or high-energy X-rays.  The
former produces a relatively weak equivalent width of the
6.4~keV line ($EW_{6.4}$) of $\sim$0.3~keV (e.g.
\cite{Ta03}), 
compared to the latter case of $EW_{6.4} \sim 1$~keV (for
the solar abundance of iron) (e.g. \cite{Mu00}). In section
3.1, we show that the K-shell lines from neutral atoms and
associated continuum (PL1) are prevailing in the
GCDX. Substituting $EW_{6.7} = 0$ in equation 3, we obtain
$EW_{6.4} = 1.4$~keV in PL1. Also, from the discussion in
section 3.1, we can say that $\mit\Gamma$ for PL1 is
$\sim$1.9.  As is noted in section 3.2, the spectral
parameters of sources 1 and 2 (table 1) include the PL1
components in the background region. However, the best-fit
$EW_{6.4} =$ 1.0--1.2~keV (table 1) and power-law index
${\mit\Gamma}=1.8$ are almost identical to those from PL1 in
the background region.  Hence, no essential change on the
spectral parameters (table 1), other than reducing the
absolute fluxes, is present.

The spectra of source 1 and source 2 were studied with
Chandra and XMM-Newton (\cite{Yu02}, \cite{Pr03},
\cite{Yu07}). However the best-fit $EW_{6.4}$ were scattered
from observation to observation.  This may be due to GCDX
subtraction, because GCDX is variable from position to
position.  In fact, the results of the off-plane background
subtraction, where $F_{6.7}$, and hence $L2_{5-10}$, is
smaller than those in the source region, give systematically
smaller $EW_{6.4}$ compared to that of the near-by
background subtraction.  We argue that the present results
of the $EW_{6.4}$$\sim$1.0--1.2~keV are reliable, because
the GCDX is taken from a nearby background, and possible
spatial variations of the GCDX are best estimated (see
section 3.1).  The $EW_{6.4}$ value of $\sim$1.0--1.2~keV is
consistent with that irradiated by X-rays, unless the iron
abundance in the east and west fields is $\sim$4--5 times
solar.  Although we have no conclusive data for the iron
abundance in the cold cloud, the iron abundance in the GCDX
is likely to be $\sim$ 1 solar (see section 4.1). Therefore,
the observed equivalent width of the 6.4~keV line in sources
1 and 2 may favor, if not be conclusive, the origin of X-ray
irradiation, rather than electron bombarding.

The iron K-edge ($N_{\rm H}$) depth at 7.1~keV is another
key parameter to judge the origin of the 6.4~keV clumps.  If
the $N_{\rm H}$ values are far larger than that of the
Galactic absorption toward the GC, then the electron origin
may not be favored (see \cite{Ta03}).  The $N_{\rm H}$ value
is, however, sensitive to the NXBG subtraction, because it
becomes significant at high energy above $\sim$7~keV. Since
Suzaku has low and stable NXBG compared to those of Chandra
and XMM-Newton (\cite{Ko07a}), we argue that the present
result of $N_{\rm H} \sim 2 \times 10^{23}$ (assuming 1
solar abundance of iron) is more reliable. This value is
somehow larger than that toward the general GC regions (see
section 4.1), but still may not be conclusive to judge
whether the origin is X-rays or electrons.

The most direct evidence to favor the X-ray origin is the
time variability of the clumps, as was reported by
\citet{Mu07} for the sub-structures of source 1.  Also, the time
variability of Sgr~B2, the other 6.4~keV clump, was found by
\citet{Ko08}.  We further confirmed the time variability of 
source 1 from the Chandra (2002) to the Suzaku (2005)
observations with a 5-$\sigma$ confidence level. The real
scale of source 1 is a few light-years, and hence the 3-year
time variability of source 1 is possible only when the
physical information travels across the source as fast as
the speed of light, like X-ray irradiation.  This speed is
impossible by electrons and/or any finite-mass particles.

\bigskip
The authors thank all of the Suzaku team members, especially
H. Uchiyama, H. Nakajima, H. Yamaguchi, and H. Mori for
their support and useful information on the XIS
performance.  This work is supported by Grant-in-Aids from
the Ministry of Education, Culture, Sports, Science and
Technology (MEXT) of Japan, the 21st Century COE "Center for
Diversity and Universality in Physics'', Scientific Research
A (KK), Priority Research Areas in Japan ``New Development
in Black Hole Astronomy''(TGT), and Grant-in-Aid for Young
Scientists B (HM). HM is also supported by the Sumitomo
Foundation, Grant for Basic Science Research Projects,
071251, 2007. TI and YH are supported by JSPS Research
Fellowship for Young Scientists.


\begin{thebibliography}{}

\bibitem[Ezuka \& Ishida(1999)]{Ez99} Ezuka, H., \& Ishida, M.\ 1999, \apjs, 120, 277 

\bibitem[Koyama et al.(1989)]{Ko89} Koyama, K., Awaki, H., 
Kunieda, H., Takano, S., \& Tawara, Y.\ 1989, \nat, 339, 603 

\bibitem[Koyama et al.(1996)]{Ko96} Koyama, K., Maeda, Y., 
Sonobe, T., Takeshima, T., Tanaka, Y., 
\& Yamauchi, S.\ 1996, \pasj, 48, 249 

\bibitem[Koyama et al.(2007a)]{Ko07a} Koyama, K., et al.\ 
2007a, \pasj, 59, S23 

\bibitem[Koyama et al.(2007b)]{Ko07b} Koyama, K., et al.\ 
2007b, \pasj, 59, S221 

\bibitem[Koyama et al.(2007c)]{Ko07c} Koyama, K., Uchiyama, 
H., Hyodo, Y., Matsumoto, H., Tsuru, T.~G., Ozaki, M., Maeda, Y., 
\& Murakami, H.\ 2007c, \pasj, 59, 237 

\bibitem[Koyama et al.(2007d)]{Ko07d} Koyama, K., et al. 2007d, \pasj, 59, S245 

\bibitem[Koyama et al.(2008)]{Ko08} Koyama, K., Inui, T., Matsumoto, H. \& Tsuru, T. 2008, \pasj, 60, S201 

\bibitem[Mitsuda et al.(2007)]{Mi07} Mitsuda, K., et al. 2007, \pasj, 59, S1

\bibitem[Mori et al.(2008)]{Mo08} Mori, H., Tsuru, T.~G., 
Hyodo, Y., Koyama, K., \& Senda, A.\ 2008, \pasj, 60, S183

\bibitem[Muno et al.(2004)]{Mu04} Muno, M.~P., et al.\ 2004, 
\apj, 613, 326 

\bibitem[Muno et al.(2007)]{Mu07} Muno, M.~P., Baganoff, 
F.~K., Brandt, W.~N., Park, S., \& Morris, M.~R.\ 2007, \apj, 656, L69 

\bibitem[Murakami et al.(2000)]{Mu00} Murakami, H., Koyama, 
K., Sakano, M., Tsujimoto, M., \& Maeda, Y.\ 2000, \apj, 534, 283 

\bibitem[Murakami et al.(2001)]{Mu01} Murakami, H., Koyama, 
K., \& Maeda, Y.\ 2001, \apj, 558, 687 

\bibitem[Nobukawa et al.(2008)]{No08} Nobukawa, M., et al.\ 
2008, \pasj, 60, S191 

\bibitem[Predehl et al.(2003)]{Pr03} Predehl, P., 
Costantini, E., Hasinger, G., 
\& Tanaka, Y.\ 2003, Astronomische Nachrichten, 324, 73 

\bibitem[Reid(1993)]{Re93} Reid, M.~J.\ 1993, \araa, 31, 345 

\bibitem[Revnivtsev et al.(2006)]{Re06} 
Revnivtsev, M., Sazonov, S., Gilfanov, M., Churazov, E., \& Sunyaev, R.\ 2006, \aap, 452, 169 

\bibitem[Revnivtsev \& Sazonov(2007)]{Re07a} Revnivtsev, M., \& Sazonov, S.\ 2007, \aap, 471, 159 

\bibitem[Revnivtsev et al.(2007)]{Re07b} 
Revnivtsev, M., Vikhlinin, A., \& Sazonov, S.\ 2007, \aap, 473, 857 

\bibitem[Serlemitsos et al.(2007)]{Se07} Serlemitsos, P.~J., 
et al.\ 2007, \pasj, 59, S9 

\bibitem[Tatischeff(2003)]{Ta03} Tatischeff, V.\ 2003, EAS 
Publications Series, 7, 79 (astro-ph/0208397v1)

\bibitem[Tsujimoto et~al.(2007)]{Ts07}  Tsujimoto, M., Hyodo, Y. \& Koyama, K. 2007, \pasj, 59, S229

\bibitem[Yusef-Zadeh et al.(2002)]{Yu02} Yusef-Zadeh, F., 
Law, C., \& Wardle, M.\ 2002, \apj, 568, L121 

\bibitem[Yusef-Zadeh et al.(2007)]{Yu07} Yusef-Zadeh, F., 
Muno, M., Wardle, M., \& Lis, D.~C.\ 2007, \apj, 656, 847 

\bibitem[Wang et al.(2006)]{Wan06} Wang, Q.~D., Dong, H., 
\& Lang, C.\ 2006, \mnras, 371, 38 

\bibitem[Warwick et al.(2006)]{War06} Warwick, R., Sakano, M., 
\& Decourchelle, A.\ 2006, Journal of Physics Conference Series, 54, 103 

\end{thebibliography}
\end{document}